\documentclass[aps,prl,reprint,superscriptaddress,longbibliography]{revtex4-2}

\usepackage{amsmath,amssymb,bm,bbm}
\usepackage{graphicx}
\usepackage{hyperref}
\usepackage{xcolor}
\colorlet{RED}{red}
\usepackage{microtype}
\usepackage[utf8]{inputenc}
\DeclareUnicodeCharacter{2081}{$_1$}
\DeclareUnicodeCharacter{2082}{$_2$}
\newcommand{\bk}{\mathbf{k}}
\newcommand{\bq}{\mathbf{q}}

\begin{document}

\title{Entropy-Driven Altermagnetism from Thermal Magnons}

\author{Tanaya Halder}
\affiliation{School of Physical Sciences, National Institute of Science Education and Research, Jatni 752050, India}
\affiliation{Homi Bhabha National Institute, Training School Complex, Anushakti Nagar, Mumbai 400094, India}
\author{Ashis K. Nandy}
\affiliation{School of Physical Sciences, National Institute of Science Education and Research, Jatni 752050, India}
\affiliation{Homi Bhabha National Institute, Training School Complex, Anushakti Nagar, Mumbai 400094, India}
\author{Anamitra Mukherjee}
\email{anamitra@niser.ac.in}
\affiliation{School of Physical Sciences, National Institute of Science Education and Research, Jatni 752050, India}
\affiliation{Homi Bhabha National Institute, Training School Complex, Anushakti Nagar, Mumbai 400094, India}
\date{\today}

\begin{abstract}
We identify a route to altermagnetism driven by entropy upon heating and
demonstrate it in a minimal square-lattice antiferromagnet. An exchange
modulation is entropically favored because the resulting momentum-dependent
magnon splitting increases the magnon entropy, thereby lowering the free
energy and favoring altermagnetism. We derive an instability criterion
for this exchange-modulated state, predict its temperature-driven
reentrant behavior, and show that its overlap with N\'eel order defines
a finite-temperature altermagnetic phase.
Variational spin-wave theory, spin--lattice Monte Carlo, and cluster diagonalization establish
the entropy-driven instability and its coexistence with magnetic order. The resulting state supports a transverse
spin conductivity. The mechanism provides a route to thermally induced
altermagnetism in compensated magnets coupled to sufficiently soft
physical modes.
\end{abstract}

\maketitle

\textit{Introduction.---}
Altermagnets combine compensated collinear magnetic order with
momentum-dependent spin splitting despite vanishing net magnetization
\cite{SmejkalSciAdv2020,SmejkalPRX2022Beyond,SmejkalPRX2022Landscape,
McClartyRauPRL2024,JungwirthNature2026}. Their defining symmetry differs
from that of a conventional antiferromagnet: opposite-spin sublattices are
related by crystal rotations rather than by translations or inversion
operations that enforce spin degeneracy. The resulting splitting changes
sign across momentum space and enables spin and transport responses without
ferromagnetic stray fields
\cite{SatoPRL2024,GhorashiPRL2024,LuPRL2024,SongNatRevMat2025}.
Momentum-resolved experiments have now established altermagnetic electronic
and magnonic signatures in several materials
\cite{KrempaskyNature2024,ReimersNatCommun2024,AminNature2024,
ZhouNature2025,YangNatCommun2025,JiangNatPhys2025,SunPRL2025,jana2026,SearsPRL2026}.
Known routes to altermagnetism generally rely on
a pre-existing or low-temperature-selected symmetry distinction
between the opposite-spin sublattices, arising from lattice, orbital,
electronic, or ferroic degrees of freedom
\cite{LeebPRL2024,FerroelectricPRL2025,MeierPRL2026,SpinOrbitalPRL2025,
AntiferroaxialPRL2026,AntiferroelasticPRB2026,ExtendedHubbardPRB2026,
JiPRB2026}.

\begin{figure*}[t]
\centering
\IfFileExists{fig1.png}{\includegraphics[width=1.0\textwidth]{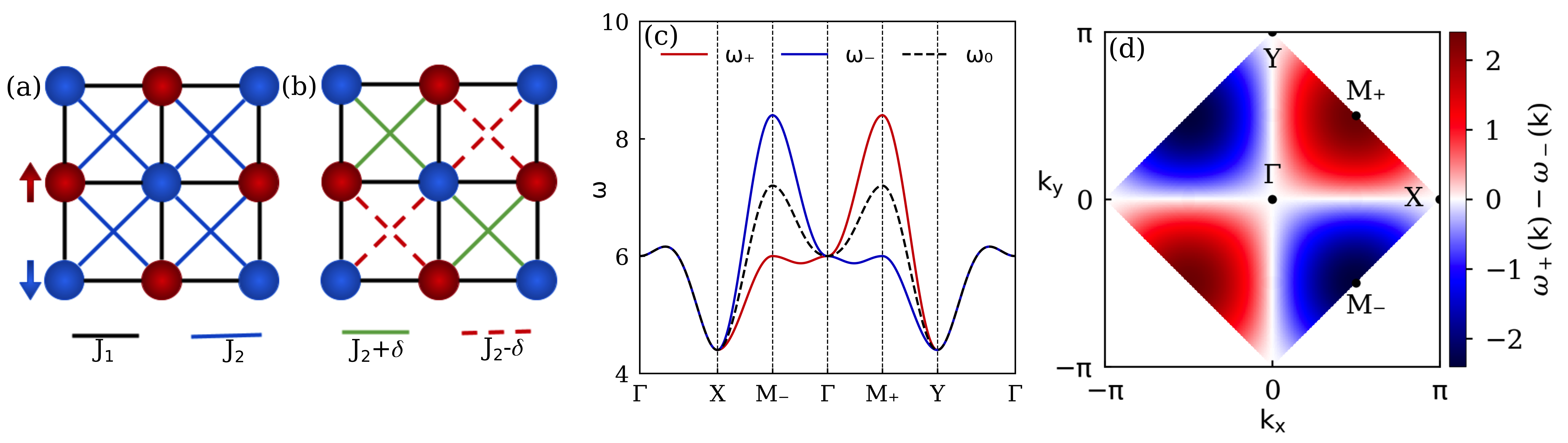}}{\fbox{\parbox[c][3.0cm][c]{1.0\columnwidth}{\centering Insert Fig.~1: variational free energy and $d_{xy}$ magnon splitting}}}
\caption{\textbf{Model and altermagnetic magnon splitting.}
(a) Parent square-lattice N\'eel antiferromagnet; red/blue denote the two
magnetic sublattices. (b) Finite-$\delta$ modulation of the diagonal
exchange. (c) Magnon bands for finite $\delta(=0.15)$, showing reversed branch
ordering along symmetry-related directions; dashed line denotes the
degenerate $\delta=0$ band $\omega_0$. (d) $d_{xy}$ sign structure of
$\omega_{+}(\mathbf{k})-\omega_{-}(\mathbf{k})$ over the magnetic Brillouin
zone (MBZ), with a representative cut. Using units with $J_1=1$, we take $J_2=0.35$, $D=0.5$, $K=0.1$, and $U=15$ to keep a gapped N\'eel background and a positive effective quartic coefficient. Spin wave results are shown for $S=2$, but can be scaled for any generic large  $S$ using the analytical formula in the text. The $K$ dependence and dimensional scales are discussed in in the text.}
\label{fig:variational}
\end{figure*}

Statistical mechanics, however, allows a qualitatively different route:
ordering one set of degrees of freedom can increase the entropy available
to another, making the more ordered state favorable upon heating. The
Pomeranchuk effect in $^3$He and inverse melting provide classic examples
of this counterintuitive competition between energy and entropy
\cite{Pomeranchuk1950,RichardsonPomeranchuk1997,Plazanet2004,SprakelEntropic2017}.
Recent work on entropic order has sharpened this principle by showing how
coupling to bosonic degrees of freedom can make an ordered configuration
entropically favorable, even in the infinite-temperature limit
\cite{HanEntropicOrder2026,HsinKobayashiEntropic2026}.
This suggests a simple mechanism for magnetic systems: a
soft symmetry-breaking exchange coordinate that costs energy at
zero temperature may become favorable at finite temperature if it
splits a thermally populated bosonic spectrum and thereby
increases its thermal entropy. Unlike conventional order-by-disorder, this
mechanism does not rely on selecting among a classically degenerate manifold;
unlike the infinite-temperature entropic-order mechanism above, it is an
intrinsically finite-temperature instability driven by thermally populated
excitations.

We demonstrate this principle in a minimal square-lattice
N\'eel antiferromagnet with antiferromagnetic nearest (NN)- and
next-nearest-neighbor (NNN) exchange. We show that, when allowed by symmetry,
magnons couple to a $d_{xy}$ modulation of the diagonal
exchange, producing the sign-changing branch splitting characteristic of
altermagnetism from an otherwise conventional antiferromagnet. Variational
spin-wave theory and spin--lattice Monte Carlo show that, upon heating, this
modulation develops while N\'eel order remains finite, generating
an intermediate altermagnetic phase before magnetic order is lost at higher
temperature. The entropy gain is explicit in the variational theory as an
enhanced magnon entropy of the modulated state, while in Monte Carlo the same
tendency appears through thermally enhanced $d_{xy}$
bond-nematic fluctuations that favor the exchange modulation.
Analytical stability analysis further shows that the thermal
softening vanishes in both the low-temperature gapped and high-temperature
paramagnetic limits, yielding a finite reentrant window in which the exchange
modulation is favored. Its overlap with the N\'eel-ordered regime defines the
thermally induced altermagnetic phase.
The resulting phase supports a transverse spin conductivity, and the mechanism
extends naturally to compensated antiferromagnets with sufficiently soft
modes promoting magnetic-exchange modulation in the appropriate altermagnetic channel.

\paragraph{Model and symmetry.---}
We consider spins of length $S$ on a square lattice with antiferromagnetic
NN and NNN exchanges, and
an easy-axis anisotropy $D$, which explicitly breaks spin-rotation
symmetry and provides a magnon gap,
as illustrated in Fig.~\ref{fig:variational}(a). The Hamiltonian is
\begin{equation}
\begin{split}
\mathcal H={}&
J_1\sum_{\langle ij\rangle}\mathbf S_i\!\cdot\!\mathbf S_j
-D\sum_i(S_i^z)^2
+\sum_{i,\mu}J_{2,i\mu}(\delta)\,
\mathbf S_i\!\cdot\!\mathbf S_{i+\mathbf d_\mu}
\\
&+N\left(\frac{K}{2}\delta^2+\frac{U}{4}\delta^4\right),
\end{split}
\label{eq:H}
\end{equation}
where $J_{2,i\mu}=J_2+\sigma_i\eta_\mu\delta$, with
$\sigma_i=\pm1$ on the two N\'eel sublattices,
$\eta_1=+1$, $\eta_2=-1$, and
$\mathbf d_{1,2}=(1,\pm1)$ denoting the two diagonal directions.
Equivalently, the modulation enters as $-\delta\mathcal O_{\rm nem}$, where
$\mathcal O_{\rm nem}
=-\sum_{i,\mu}\sigma_i\eta_\mu\,
\mathbf S_i\!\cdot\!\mathbf S_{i+\mathbf d_\mu}$
is the bond-nematic operator conjugate to the exchange coordinate $\delta$.
Since $\mathcal O_{\rm nem}$ changes sign under a $C_4$ rotation, finite
$\delta$ selects one of two $C_4$-related diagonal-exchange patterns.
The opposite signs on the two diagonal directions generate a momentum-space
form factor proportional to $\sin k_x\sin k_y$, identifying the bond-nematic
exchange modulation as $d_{xy}$
[Fig.~\ref{fig:variational}(b); see \cite{sm}, Sec.~S1].
Microscopically, such an exchange coordinate can arise from the sensitivity
of superexchange to orbital and lattice distortions, as in
Kugel--Khomskii-type exchange\cite{Kugel1982}, with a soft phonon mode modulating the
diagonal exchange couplings. The $K,U>0$ terms favor $\delta=0$ in the absence of thermal magnetic
fluctuations. We set $J_1=1$ throughout and quote all parameters in the
corresponding dimensionless units; the chosen parameters keep the collinear
N\'eel state locally stable over the relevant range of $\delta$.

For $\delta=0$, the N\'eel state is invariant under a one-site translation followed by spin reversal. Finite $\delta$ makes the two diagonal-exchange environments inequivalent, breaking this translation--spin-reversal symmetry while preserving the
plaquette-centered $C_4$ rotation combined with spin reversal required for altermagnetism. The state therefore remains compensated but acquires altermagnetic symmetry.

We use representative parameters on the N\'eel side of the unmodulated exchange model: $J_2/J_1=0.35$ and easy-axis anisotropy $D/J_1=0.5$, which opens a magnon gap and allows finite-temperature order in 2D. The quartic coefficient is set to yield a continuous exchange-modulation mode onset while keeping the equilibrium modulation in the locally stable spin-wave regime. We examine the $K$ dependence and pertinent phase diagram later in the paper.

Linear spin-wave theory (LSWT) about the N\'eel state \cite{hp} yields two
opposite-spin magnon branches (\cite{sm}, Sec.~S2),
$\omega_\pm(\mathbf k,\delta)=R_{\mathbf k}\pm\lambda_{\mathbf k}\delta$,
with $R_{\mathbf k}=\sqrt{\Omega_{\mathbf k}^2-B_{\mathbf k}^2}$,
$\Omega_{\mathbf k}=4J_1S+2DS-4J_2S(1-\cos k_x\cos k_y)$,
$B_{\mathbf k}=2J_1S(\cos k_x+\cos k_y)$, and
$\lambda_{\mathbf k}=4S\sin k_x\sin k_y$.
The modulation therefore produces
$\omega_+-\omega_-=8S\delta\sin k_x\sin k_y$, a $d_{xy}$ splitting that
vanishes on the coordinate axes and changes sign under
$(k_x,k_y)\!\rightarrow\!(k_x,-k_y)$ and under $C_4$.
Accordingly, the two spin branches reverse their ordering along
$\Gamma\!\rightarrow\!M_+$ and $\Gamma\!\rightarrow\!M_-$
[Fig.~\ref{fig:variational}(c,d)].
Finite $\delta$ (value determined by variational minimization discussed in the next paragraph) thus converts the spin-degenerate antiferromagnetic magnons
into the momentum-dependent, sign-changing spectrum characteristic of
altermagnetism.

\begin{figure*}[t]
\centering
\IfFileExists{fig2.png}{\includegraphics[width=1.0\textwidth]{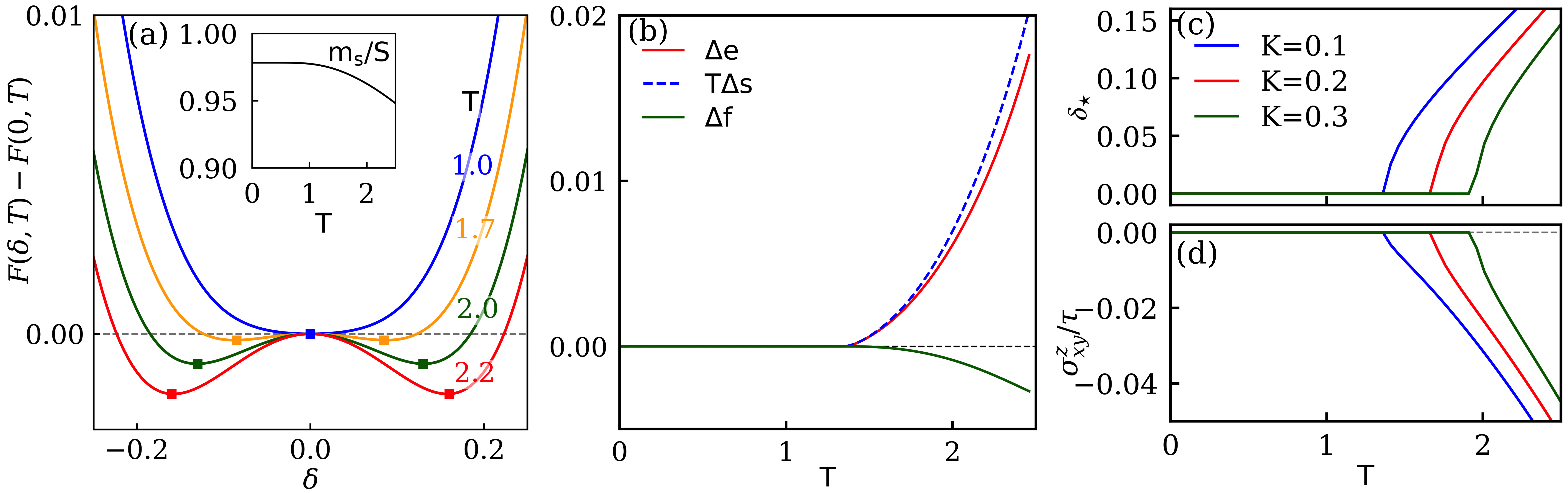}}{\fbox{\parbox[c][3.0cm][c]{1.0\textwidth}{\centering Insert final three-panel MC/finite-size/transport figure}}}
\caption{\textbf{Entropy-driven altermagnetic instability.}
(a) Variational free-energy difference
$\Delta f(\delta,T)=F(\delta,T)-F(0,T)$ at representative temperatures.
Upon heating, the minimum shifts from $\delta=0$ to symmetry-related
finite $\pm\delta_\star$; inset: spin-wave-renormalized staggered moment
evaluated at $\delta_\star(T)$.
(b) Free-energy balance at $\delta_\star(T)$: energetic cost $\Delta e$,
entropic gain $T\Delta s$, and $\Delta f=\Delta e-T\Delta s$.
Panels (a,b) use the same parameters as Fig.~\ref{fig:variational}.
(c) Equilibrium modulation $\delta_\star$ and 
(d) Corresponding transverse spin conductivity per relaxation time,
$\left\langle
\sigma_{xy}^{z}\right\rangle/{\tau}
$, versus temperature, is expressed in units of
$k_{\mathrm B}J_1/\hbar$, for the indicated values of $K$, with all other
parameters unchanged.}
\label{fig2}
\end{figure*}

\paragraph{Entropy-driven instability.---}
At $T=0$, the bare exchange-mode energy favors $\delta=0$. At finite
temperature, thermally occupied magnons provide an additional
$\delta$-dependent contribution to the free energy. For the staggered
$d_{xy}$ modulation, both the classical N\'eel energy and the zero-point
energy are independent of $\delta$: the $+\delta$ and $-\delta$ bond
contributions cancel in the former, while
$\omega_+(\bk,\delta)+\omega_-(\bk,\delta)=2R_{\bk}$ in the latter.
The relevant free energy  therefore contains only the elastic energy  preferring $\delta=0$, and the magnon contributions integrated over the full magnetic Brillouin zone (MBZ):
\begin{equation}
F(\delta,T)=
\frac{K}{2}\delta^2+\frac{U}{4}\delta^4+
T\sum_{s=\pm}\int_{\rm MBZ}\!\frac{d^2k}{(2\pi)^2}
\ln\!\left(1-e^{-\omega_s/T}\right).
\label{eq:fullfree}
\end{equation}
Because $\omega_\pm=R_{\bk}\pm\lambda_{\bk}\delta$, the linear thermal
corrections cancel, whereas the leading quadratic correction is negative.
Expanding about $\delta=0$ therefore renormalizes the exchange stiffness as
$K_{\rm eff}(T)=K-\chi_{\rm nem}^{\rm SW}(T)$, where
$\chi_{\rm nem}^{\rm SW}$ is the thermal $d_{xy}$ bond-nematic
susceptibility. Physically, these thermally activated
nematic fluctuations reflect the entropy gained by splitting the populated magnon branches; See \cite{sm}, Sec.~S3 for details. Once $\chi_{\rm nem}^{\rm SW}(T)>K$, the unmodulated state becomes unstable to symmetry-related minima at $\pm\delta$. In the regime studied, this instability occurs well below $T_{\rm N}$ and
away from the magnetic critical region, so the N\'eel background remains ordered and $\chi_{\rm nem}^{\rm SW}$ remains finite. A sufficiently positive quartic coefficient then stabilizes the finite-$\delta$ state.

The numerical minimization of Eq.~\eqref{eq:fullfree} directly demonstrates
this instability. As shown in Fig.~\ref{fig2}(a), the free-energy minimum
lies at $\delta=0$ at low temperature and splits into symmetry-related minima
at finite $\pm\delta_\star$ upon heating. The inset shows that the
spin-wave-renormalized staggered moment remains finite across this onset,
placing the instability within the N\'eel phase; for details see
\cite{sm}, Sec.~S3]. Its entropic origin is made
explicit in Fig.~\ref{fig2}(b): at the equilibrium $\delta_\star(T)$, the
entropic gain $T\Delta s$ overcomes the energetic cost $\Delta e$, yielding
$\Delta f=\Delta e-T\Delta s<0$. Increasing the bare stiffness $K$ shifts the
onset of finite $\delta_\star$ to higher temperatures
[Fig.~\ref{fig2}(c)], consistent with a harder exchange coordinate requiring
stronger thermal fluctuations to become unstable. Thus, the exchange
modulation is stabilized thermally while magnetic order persists, producing
the altermagnetic phase.

\textit{Transverse spin transport.---}
The thermally generated altermagnetic state acquires a transverse spin
response to a longitudinal temperature gradient through its momentum-dependent
magnon splitting. For $\delta=0$, the contributions
from the two opposite-spin branches cancel, whereas finite $\delta$ removes
this cancellation. Symmetry requires
$\sigma_{xx}^{z}=\sigma_{yy}^{z}=0$ and
$\sigma_{xy}^{z}=\sigma_{yx}^{z}$. Within a constant-relaxation-time
Boltzmann treatment \cite{PhysRevB.108.L180401},
\begin{equation}
\frac{\sigma_{xy}^{z}}{\tau}
=
\sum_{s=\pm1}s
\int_{\rm MBZ}\frac{d^2k}{(2\pi)^2}\,
v_{s,x}v_{s,y}
\frac{\partial n_B(\omega_s)}{\partial T},
\label{eq:sigmafull}
\end{equation}
where $v_{s,a}=\partial_{k_a}\omega_s$, employing $\hbar=K_B=1$.
The response is odd under reversal of the altermagnetic domain,
$\sigma_{xy}^{z}(-\delta)=-\sigma_{xy}^{z}(\delta)$.

At low temperature, transport is dominated by thermally activated magnons
near $X=(\pi,0)$, where the $d_{xy}$ splitting vanishes but modifies the
surrounding magnon velocities. Expanding about $X$ gives
$\sigma_{xy}^{z}/\tau\simeq
-(\delta/\pi J_2)(\Delta_X+2T)e^{-\Delta_X/T}$, with
$\Delta_X=4J_1S+2DS-8J_2S$; the derivation is given in
\cite{sm}, Sec.~S4. The exponential factor reflects activation across the
magnon gap, while the factor of $\delta$ gives the sign reversal under
$\delta\rightarrow-\delta$.

Using the equilibrium $\delta_\star(T)$ obtained from Eq.~\eqref{eq:fullfree}, the full MBZ evaluation of Eq.~\eqref{eq:sigmafull} gives the conductivity per relaxation time, shown in Fig.~\ref{fig2}(d). It develops together with the finite exchange modulation in Fig.~\ref{fig2}(c), with both the onset and magnitude controlled by the bare stiffness $K$. 

\begin{figure}[t]
\centering
\IfFileExists{fig3.png}{\includegraphics[width=0.48\textwidth]{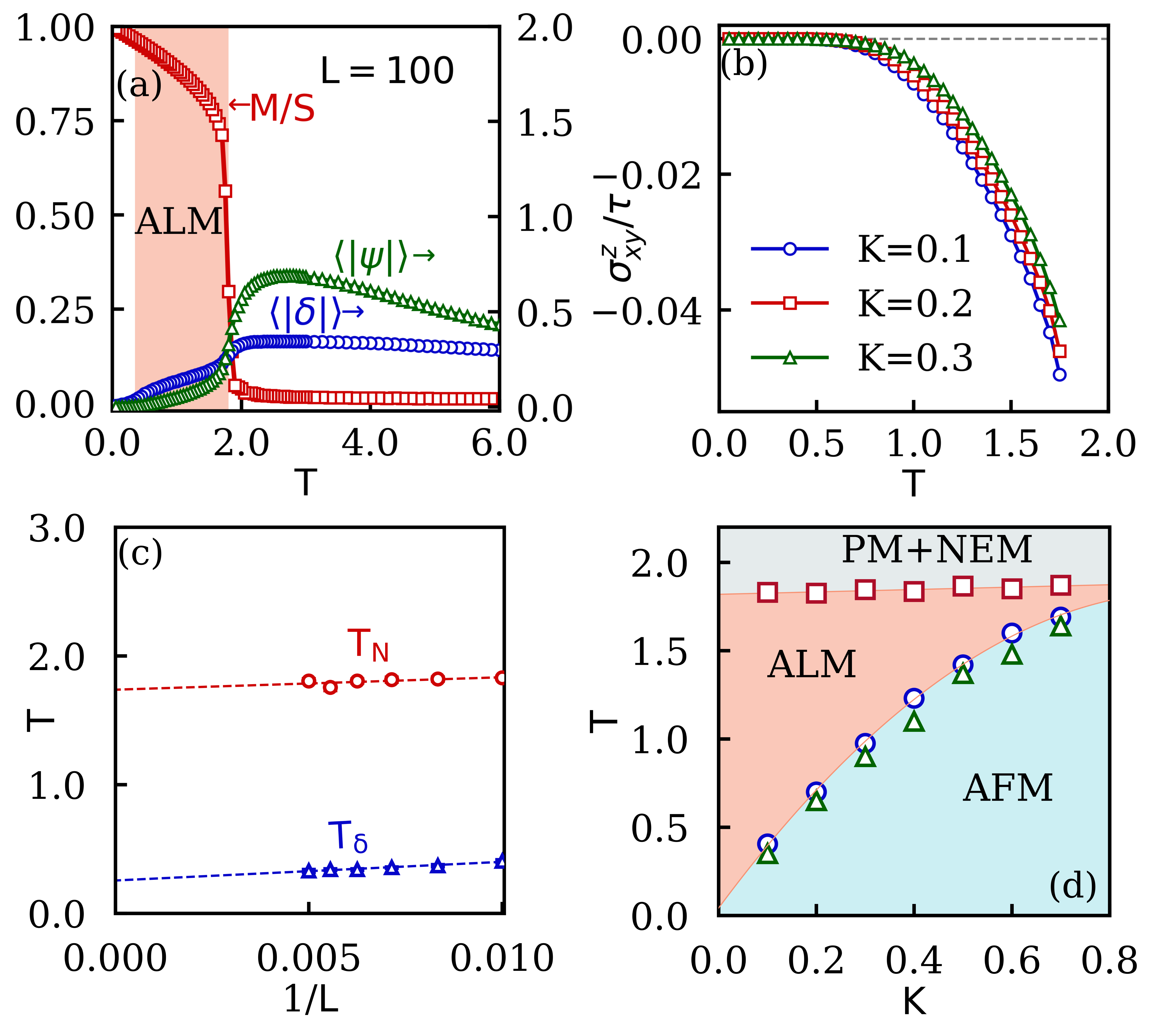}}{\fbox{\parbox[c][4.0cm][c]{0.4\textwidth}{\centering Insert final three-panel MC/finite-size/transport figure}}}
\caption{\textbf{Monte Carlo phase diagram \& spin transport.}
(a) N\'eel order $M/S$, exchange modulation $\langle|\delta|\rangle$,
and bond-nematic order $\langle|\psi|\rangle$ for $L=100$.
(b) Transverse spin conductivity per relaxation time,
$\sigma_{xy}^{z}/\tau$, for the indicated $K$.
(c) Finite-size transition temperatures versus $1/L$;
dashed lines indicate thermodynamic-limit extrapolations.
(d) $K$--$T$ phase diagram: Altermagnet (ALM) lies between low-$T$ antiferromagnet (AFM) and
high-$T$ nematic paramagnet (PM+Nem); the higher-$T$ PM phase is
not shown. Symbols are defined in the text.
Panels (a,c) use $K=0.1$; other parameters are as in
Fig.~\ref{fig:variational}.}
\label{fig:f3}
\end{figure}

\textit{Thermodynamics and phase structure.---}
To test the instability beyond LSWT, we perform classical spin--lattice
Monte Carlo directly on Eq.~\eqref{eq:H}, sampling both the spins and the
signed global exchange coordinate $\delta$. As discussed above, $\delta$
couples linearly to the $d_{xy}$ bond-nematic order parameter $\psi$.
In the classical theory, integrating out the spin fluctuations produces the
same thermal softening of the exchange coordinate $\delta$ through the $d_{xy}$ bond-nematic response as discussed below.

We simulate lattices up to $200\times200$ and monitor the N\'eel order $M$,
the bond-nematic order $\langle|\psi|\rangle$, and the exchange modulation
$\langle|\delta|\rangle$; details of the Monte Carlo protocol are given in
\cite{sm}, Sec.~S5. These quantities provide the thermodynamic diagnostics
shown in Fig.~\ref{fig:f3}.
Figure~\ref{fig:f3}(a) shows the thermal evolution of the coupled
spin--exchange system. On heating through $T_\delta$,
$\langle|\delta|\rangle$ and the conjugate bond-nematic order
$\langle|\psi|\rangle$ develop together while the N\'eel order remains
finite, signaling spontaneous $C_4$ breaking within the magnetically ordered
phase. We locate $T_\delta(L)$ from the peak of $\chi_\delta$ and $T_N(L)$
from the magnetic susceptibility. Both admit leading $1/L$ finite-size
shifts, see \cite{sm} Sec. S5: for the coherent exchange mode the scaling variable is
$(T-T_\delta)N^{1/2}$ with $N=L^2$, while the easy-axis magnetic transition
has $\nu=1$. Their extrapolations [Fig.~\ref{fig:f3}(c)] yield
$T_\delta<T_N$, establishing an intermediate altermagnetic regime. The
exchange modulation persists above $T_N$ over the temperature range shown,
but this regime is not altermagnetic because magnetic long-range order is
absent.

The corresponding transverse spin conductivity is shown in
Fig.~\ref{fig:f3}(b). For each temperature, Eq.~\eqref{eq:sigmafull} is
evaluated for the thermally sampled values of $|\delta|$ and averaged over
their Monte Carlo distribution; the spin-wave spectrum supplies the transport
kernel, while Monte Carlo supplies the equilibrium fluctuations of the
exchange coordinate. We restrict this construction to $T<T_N$, where the
N\'eel-state magnon description remains applicable. Increasing $K$ shifts the
onset of the response to higher temperature, consistent with the larger
thermal softening required to stabilize finite $\delta$.

\textit{Finite-temperature window.---}
In the classical Monte Carlo description, integrating out the spins and expanding the free energy in the exchange coordinate gives
$K_{\rm eff}(T)=K-\chi_{\rm nem}^{(0)}(T)$, where
$\chi_{\rm nem}^{(0)}$ is the $d_{xy}$ bond-nematic susceptibility of
the unmodulated ($\delta=0$) spin system (\cite{sm}, Sec.~S6).  This is the classical counterpart of the spin-wave
softening governed by $\chi_{\rm nem}^{\rm SW}(T)$.
Figure~\ref{fig:f3}(d) maps the altermagnetic 
(ALM) region between $T_\delta$ 
(circles) estimates from $\chi_\delta$ and AFM $T_N$ (squares) in the $K$--$T$ plane. Comparing the onset $T_\delta$ (circles) with those obtained from the crossings of
$K$ and $\chi_{\rm nem}^{(0)}(T)$ (triangles)  provides a numerical check of the exchange-softening criterion.

This criterion helps explain the lower onset temperatures in Monte
Carlo than in the variational calculation, and why the finite T softening is confined to a finite temperature window. For gapped quantum magnons,
$n_B(R_{\mathbf k})\sim e^{-R_{\mathbf k}/T}$ exponentially suppresses
$\chi_{\rm nem}^{\rm SW}$ at low temperature. The classical
Rayleigh--Jeans occupation $n_B(R_{\mathbf k})\simeq T/R_{\mathbf k}$
instead gives $\chi_{\rm nem}^{(0)}\propto T$, favoring softening at
lower temperatures than in the spin-wave case. Further, since the classical susceptibility vanishes
as $1/T$ at high temperature, the condition
$\chi_{\rm nem}^{(0)}(T)>K$ naturally \textit{confines the softening to a finite-temperature
window} for $K>0$; its overlap with N\'eel order defines the
altermagnetic phase. Increasing $K$ narrows this overlap, which is
expected to disappear when softening no longer occurs within the
ordered regime. Conversely, $T_\delta\to0$ as $K\to0^+$.
Finally, small-cluster exact diagonalization also finds thermally enhanced
nematic susceptibility and negative effective stiffness, providing
a finite-cluster quantum check of the mechanism
(\cite{sm}, Sec.~S7).

\textit{Routes for experimental realization.---}
The mechanism requires a compensated antiferromagnet, a sufficiently soft
exchange-active coordinate in an altermagnetic symmetry channel, and
thermally populated magnetic excitations whose entropy favors the modulated
state. A direct material realization of this combination has not, to our
knowledge, been established, although its individual ingredients have clear
precedents. Entropy can stabilize finite-temperature magnetic order or bond
modulation in spin systems
\cite{SirkerKhaliullin2003,SaglamNatPhys2022}, while strong
temperature-dependent exchange driven by ligand or molecular motion has been observed in CsO$_2$ \cite{KlanjsekPRL2015}.
The relation between the exchange modulation $\delta$ and a symmetry-compatible
structural displacement, together with illustrative displacement and
magnon-splitting scales, is discussed in \cite{sm} Sec. S1. 
More broadly, such
an exchange-active phonon could also be driven coherently: resonant phonon
excitation is known to modify magnetic exchange on ultrafast timescales
\cite{AfanasievNatMater2021}, offering a nonequilibrium route to transiently
generate the same altermagnetic exchange pattern.

\textit{Conclusion.---}
We have shown that thermal fluctuations can generate, rather than merely
destroy, the symmetry breaking required for altermagnetism. The key principle
is that a soft exchange coordinate can become favorable upon heating when the
resulting symmetry breaking increases the entropy of thermally populated
magnetic excitations. Because this entropic softening vanishes in both the
low- and high-temperature limits, the instability naturally occupies a finite
thermal window. In our minimal antiferromagnet, this window overlaps the
N\'eel phase, producing a thermally induced altermagnetic state with
sign-changing magnon splitting and a transverse spin response. More broadly,
the same principle suggests a route to finite-temperature magnetic phases
whenever a sufficiently soft exchange-active degree of freedom couples to an
excitation spectrum that gains entropy upon symmetry breaking.

\bibliography{entropy_driven_altermagnetism_revised}

@article{SmejkalSciAdv2020,
  author  = {L. {\v{S}}mejkal and R. Gonz{\'a}lez-Hern{\'a}ndez and T. Jungwirth and J. Sinova},
  title   = {Crystal time-reversal symmetry breaking and spontaneous Hall effect in collinear antiferromagnets},
  journal = {Science Advances},
  volume  = {6},
  pages   = {eaaz8809},
  year    = {2020},
  doi     = {10.1126/sciadv.aaz8809}
}

@article{KlanjsekPRL2015,
  author  = {Klanj{\v{s}}ek, M. and Ar{\v{c}}on, D. and Sans, A. and
             Adler, P. and Jansen, M. and Felser, C.},
  title   = {Phonon-Modulated Magnetic Interactions and Spin
             Tomonaga--Luttinger Liquid in the $p$-Orbital
             Antiferromagnet CsO$_2$},
  journal = {Phys. Rev. Lett.},
  volume  = {115},
  pages   = {057205},
  year    = {2015},
  doi     = {10.1103/PhysRevLett.115.057205}
}

@article{SirkerKhaliullin2003,
  author  = {Sirker, J. and Khaliullin, G.},
  title   = {Entropy Driven Dimerization in a One-Dimensional
             Spin-Orbital Model},
  journal = {Phys. Rev. B},
  volume  = {67},
  pages   = {100408},
  year    = {2003},
  doi     = {10.1103/PhysRevB.67.100408}
}

@article{SaglamNatPhys2022,
  author  = {Saglam, H. and Duzgun, A. and Kargioti, A. and Harle, N. and
             Zhang, X. and Bingham, N. S. and Lao, Y. and Gilbert, I. and
             Sklenar, J. and Watts, J. D. and Ramberger, J. and Bromley, D.
             and Chopdekar, R. V. and O'Brien, L. and Leighton, C. and
             Nisoli, C. and Schiffer, P.},
  title   = {Entropy-driven order in an array of nanomagnets},
  journal = {Nat. Phys.},
  volume  = {18},
  pages   = {706--712},
  year    = {2022},
  doi     = {10.1038/s41567-022-01555-6}
}

@article{LarkinPikin1969,
  author  = {Larkin, A. I. and Pikin, S. A.},
  title   = {Phase transitions of the first order but nearly of the second},
  journal = {Sov. Phys. JETP},
  volume  = {29},
  number  = {5},
  pages   = {891--896},
  year    = {1969},
  url     = {https://www.jetp.ras.ru/cgi-bin/dn/e_029_05_0891.pdf}
}

@article{ChandraPRResearch2020,
  author  = {Chandra, Premala and Coleman, Piers and
             Continentino, Mucio A. and Lonzarich, Gilbert G.},
  title   = {Quantum annealed criticality: A scaling description},
  journal = {Phys. Rev. Research},
  volume  = {2},
  number  = {4},
  pages   = {043440},
  year    = {2020},
  doi     = {10.1103/PhysRevResearch.2.043440}
}

@article{AfanasievNatMater2021,
  author  = {Afanasiev, D. and Hortensius, J. R. and Ivanov, B. A. and
             Sasani, A. and Bousquet, E. and Blanter, Y. M. and
             Mikhaylovskiy, R. V. and Kimel, A. V. and Caviglia, A. D.},
  title   = {Ultrafast control of magnetic interactions via light-driven
             phonons},
  journal = {Nat. Mater.},
  volume  = {20},
  pages   = {607--611},
  year    = {2021},
  doi     = {10.1038/s41563-021-00922-7}
}

@article{SmejkalPRX2022Beyond,
  author  = {L. {\v{S}}mejkal and J. Sinova and T. Jungwirth},
  title   = {Beyond Conventional Ferromagnetism and Antiferromagnetism: A Phase with Nonrelativistic Spin and Crystal Rotation Symmetry},
  journal = {Physical Review X},
  volume  = {12},
  pages   = {031042},
  year    = {2022},
  doi     = {10.1103/PhysRevX.12.031042}
}

@article{SmejkalPRX2022Landscape,
  author  = {L. {\v{S}}mejkal and J. Sinova and T. Jungwirth},
  title   = {Emerging Research Landscape of Altermagnetism},
  journal = {Physical Review X},
  volume  = {12},
  pages   = {040501},
  year    = {2022},
  doi     = {10.1103/PhysRevX.12.040501}
}

@article{McClartyRauPRL2024,
  author  = {P. A. McClarty and J. G. Rau},
  title   = {Landau Theory of Altermagnetism},
  journal = {Physical Review Letters},
  volume  = {132},
  pages   = {176702},
  year    = {2024},
  doi     = {10.1103/PhysRevLett.132.176702}
}

@article{LeebPRL2024,
  author  = {V. Leeb and A. Mook and L. {\v{S}}mejkal and J. Knolle},
  title   = {Spontaneous Formation of Altermagnetism from Orbital Ordering},
  journal = {Physical Review Letters},
  volume  = {132},
  pages   = {236701},
  year    = {2024},
  doi     = {10.1103/PhysRevLett.132.236701}
}

@article{SatoPRL2024,
  author  = {T. Sato and S. Haddad and I. C. Fulga and F. F. Assaad and J. van den Brink},
  title   = {Altermagnetic Anomalous Hall Effect Emerging from Electronic Correlations},
  journal = {Physical Review Letters},
  volume  = {133},
  pages   = {086503},
  year    = {2024},
  doi     = {10.1103/PhysRevLett.133.086503}
}

@article{Kugel1982,
  author    = {Kugel', K. I. and Khomskii, D. I.},
  title     = {The Jahn-Teller effect and magnetism: transition metal compounds},
  journal   = {Soviet Physics Uspekhi},
  volume    = {25},
  number    = {4},
  pages     = {231--256},
  year      = {1982},
  doi       = {10.1070/PU1982v025n04ABEH004537},
  url       = {https://iop.org}
}

@article{KrempaskyNature2024,
  author  = {J. Krempask{\'y} and L. {\v{S}}mejkal and S. W. D'Souza and M. Hajlaoui and G. Springholz and K. Uhl{\'i}{\v{r}}ov{\'a} and others},
  title   = {Altermagnetic lifting of Kramers spin degeneracy},
  journal = {Nature},
  volume  = {626},
  pages   = {517--522},
  year    = {2024},
  doi     = {10.1038/s41586-023-06907-7}
}

@article{ReimersNatCommun2024,
  author  = {S. Reimers and L. Odenbreit and L. {\v{S}}mejkal and V. N. Strocov and P. Constantinou and A. B. Hellenes and others},
  title   = {Direct observation of altermagnetic band splitting in CrSb thin films},
  journal = {Nature Communications},
  volume  = {15},
  pages   = {2116},
  year    = {2024},
  doi     = {10.1038/s41467-024-46476-5}
}

@article{AminNature2024,
  author  = {O. J. Amin and A. Dal Din and E. Golias and Y. Niu and A. Zakharov and S. C. Fromage and others},
  title   = {Nanoscale imaging and control of altermagnetism in MnTe},
  journal = {Nature},
  volume  = {636},
  pages   = {348--353},
  year    = {2024},
  doi     = {10.1038/s41586-024-08234-x}
}

@article{GhorashiPRL2024,
  author  = {S. A. A. Ghorashi and T. L. Hughes and J. Cano},
  title   = {Altermagnetic Routes to Majorana Modes in Zero Net Magnetization},
  journal = {Physical Review Letters},
  volume  = {133},
  pages   = {106601},
  year    = {2024},
  doi     = {10.1103/PhysRevLett.133.106601}
}

@article{LuPRL2024,
  author  = {B. Lu and K. Maeda and H. Ito and K. Yada and Y. Tanaka},
  title   = {$\phi$ Josephson Junction Induced by Altermagnetism},
  journal = {Physical Review Letters},
  volume  = {133},
  pages   = {226002},
  year    = {2024},
  doi     = {10.1103/PhysRevLett.133.226002}
}

@article{FerroelectricPRL2025,
  author  = {X. Chen and others},
  title   = {Ferroelectric Switchable Altermagnetism},
  journal = {Physical Review Letters},
  volume  = {134},
  pages   = {106802},
  year    = {2025},
  doi     = {10.1103/PhysRevLett.134.106802}
}

@article{ZhouNature2025,
  author  = {Z. Zhou and X. Cheng and M. Hu and R. Chu and H. Bai and L. Han and J. Liu and F. Pan and others},
  title   = {Manipulation of the altermagnetic order in CrSb via crystal symmetry},
  journal = {Nature},
  volume  = {638},
  pages   = {645--650},
  year    = {2025},
  doi     = {10.1038/s41586-024-08436-3}
}

@article{YangNatCommun2025,
  author  = {G. Yang and Z. Li and S. Yang and J. Li and H. Zheng and W. Zhu and others},
  title   = {Three-dimensional mapping of the altermagnetic spin splitting in CrSb},
  journal = {Nature Communications},
  volume  = {16},
  year    = {2025},
  doi     = {10.1038/s41467-025-56647-7}
}

@article{JiangNatPhys2025,
  author  = {B. Jiang and M. Hu and J. Bai and others},
  title   = {A metallic room-temperature $d$-wave altermagnet},
  journal = {Nature Physics},
  volume  = {21},
  pages   = {754--759},
  year    = {2025},
  doi     = {10.1038/s41567-025-02822-y}
}

@article{SongNatRevMat2025,
  author  = {C. Song and H. Bai and Z. Zhou and L. Han and H. Reichlova and J. H. Dil and J. Liu and X. Chen and others},
  title   = {Altermagnets as a new class of functional materials},
  journal = {Nature Reviews Materials},
  volume  = {10},
  pages   = {473--485},
  year    = {2025},
  doi     = {10.1038/s41578-025-00779-1}
}

@article{SunPRL2025,
  author  = {Q. Sun and J. Guo and D. Wang and D. L. Abernathy and W. Tian and C. Li},
  title   = {Observation of Chiral Magnon Band Splitting in Altermagnetic Hematite},
  journal = {Physical Review Letters},
  volume  = {135},
  pages   = {186703},
  year    = {2025},
  doi     = {10.1103/7yhz-jptc}
}

@article{SpinOrbitalPRL2025,
  author  = {Y. Zhang and others},
  title   = {Spin-Orbital Altermagnetism},
  journal = {Physical Review Letters},
  volume  = {135},
  pages   = {176705},
  year    = {2025},
  doi     = {10.1103/cjzw-j4v7}
}

@article{MeierPRL2026,
  author  = {Q. N. Meier and A. Carta and C. Ederer and A. Cano},
  title   = {Net and Compensated Altermagnetism from Staggered Orbital Order: Layer-Dependent Spin Splitting in Sr$_{n+1}$Cr$_n$O$_{3n+1}$},
  journal = {Physical Review Letters},
  volume  = {136},
  pages   = {116705},
  year    = {2026},
  doi     = {10.1103/mm8t-82q4}
}

@article{SearsPRL2026,
  author  = {J. Sears and V. O. Garlea and D. Lederman and J. M. Tranquada and I. A. Zaliznyak},
  title   = {Altermagnetic and Dipolar Splitting of Magnons in FeF$_2$},
  journal = {Physical Review Letters},
  volume  = {136},
  pages   = {226701},
  year    = {2026},
  doi     = {10.1103/g6dt-rf8c}
}

@article{AntiferroaxialPRL2026,
  author  = {Y. Liu and C.-C. Liu},
  title   = {Antiferroaxial Altermagnetism},
  journal = {Physical Review Letters},
  volume  = {136},
  pages   = {256709},
  year    = {2026},
  doi     = {10.1103/21z4-c9p2}
}

@article{AntiferroelasticPRB2026,
  author  = {F. Han and S. Tang and J.-D. Sun and X.-G. Zhao and H. Xu},
  title   = {Strain-driven antiferroelastic switching of altermagnetism},
  journal = {Physical Review B},
  volume  = {113},
  pages   = {214457},
  year    = {2026},
  doi     = {10.1103/t7hz-hs8v}
}

@article{ExtendedHubbardPRB2026,
  author  = {Y. Liu and others},
  title   = {Spontaneous emergence of altermagnetism in the single-orbital extended Hubbard model},
  journal = {Physical Review B},
  volume  = {113},
  pages   = {245117},
  year    = {2026},
  doi     = {10.1103/k5vw-c9ks}
}

@article{JiPRB2026,
  author  = {C. Ji and K. Chen and Q. Chen and C.-K. Duan},
  title   = {Spin-orbital altermagnetism in strongly-correlated fluoroperovskites},
  journal = {Physical Review B},
  volume  = {113},
  pages   = {245102},
  year    = {2026},
  doi     = {10.1103/txt9-t77j}
}

@article{JungwirthNature2026,
  author  = {T. Jungwirth and J. Sinova and R. M. Fernandes and others},
  title   = {Symmetry, microscopy and spectroscopy signatures of altermagnetism},
  journal = {Nature},
  volume  = {649},
  pages   = {837--847},
  year    = {2026},
  doi     = {10.1038/s41586-025-09883-2}
}

@article{HanEntropicOrder2026,
  author  = {Han, Yiqiu and Huang, Xiaoyang and Komargodski, Zohar
             and Lucas, Andrew and Popov, Fedor K.},
  title   = {Entropic Order},
  journal = {Nature Communications},
  volume  = {17},
  pages   = {87},
  year    = {2026},
  doi     = {10.1038/s41467-025-66797-3}
}

@misc{HsinKobayashiEntropic2026,
  author        = {Hsin, Po-Shen and Kobayashi, Ryohei},
  title         = {Exploring Entropic Orders: High Temperature Continuous
                   Symmetry Breaking, Chiral Topological States and Local
                   Commuting Projector Models},
  year          = {2026},
  eprint        = {2604.18694},
  archivePrefix = {arXiv},
  primaryClass  = {cond-mat.str-el}
}

@article{Pomeranchuk1950,
  author  = {Pomeranchuk, I. Ya.},
  title   = {On the Theory of Liquid Helium-3},
  journal = {Zh. Eksp. Teor. Fiz.},
  volume  = {20},
  pages   = {919},
  year    = {1950}
}

@article{RichardsonPomeranchuk1997,
  author  = {Richardson, Robert C.},
  title   = {The Pomeranchuk Effect},
  journal = {Reviews of Modern Physics},
  volume  = {69},
  pages   = {683--690},
  year    = {1997},
  doi     = {10.1103/RevModPhys.69.683}
}

@article{Plazanet2004,
  author  = {Plazanet, M. and Floare, C. and Johnson, M. R.
             and Schweins, R. and Trommsdorff, H. P.},
  title   = {Freezing on Heating of Liquid Solutions},
  journal = {The Journal of Chemical Physics},
  volume  = {121},
  pages   = {5031--5034},
  year    = {2004},
  doi     = {10.1063/1.1794652}
}

@article{SprakelEntropic2017,
  author  = {Sprakel, Joris and Zaccone, Alessio and Spaepen, Frans
             and Schall, Peter and Weitz, David A.},
  title   = {Direct Observation of Entropic Stabilization of bcc Crystals
             Near Melting},
  journal = {Physical Review Letters},
  volume  = {118},
  pages   = {088003},
  year    = {2017},
  doi     = {10.1103/PhysRevLett.118.088003}
}

@misc{jana2026,
      title={Correlation-Driven Orbital Order Realizes 2D Metallic Altermagnetism}, 
      author={Nirmalya Jana and Atasi Chakraborty and Anamitra Mukherjee and Amit Agarwal},
      year={2026},
      eprint={2603.25426},
      archivePrefix={arXiv},
      primaryClass={cond-mat.mes-hall},
      url={https://arxiv.org/abs/2603.25426}, 
}

@article{PhysRevB.108.L180401,
  title = {Efficient spin Seebeck and spin Nernst effects of magnons in altermagnets},
  author = {Cui, Qirui and Zeng, Bowen and Cui, Ping and Yu, Tao and Yang, Hongxin},
  journal = {Phys. Rev. B},
  volume = {108},
  issue = {18},
  pages = {L180401},
  numpages = {7},
  year = {2023},
  month = {Nov},
  publisher = {American Physical Society},
  doi = {10.1103/PhysRevB.108.L180401},
  url = {https://link.aps.org/doi/10.1103/PhysRevB.108.L180401}
}

@article{hp, 
  year     = {1940}, 
  title    = {Field Dependence of the Intrinsic Domain Magnetization of a Ferromagnet}, 
  author   = {Holstein, T and Primakoff, H}, 
  journal  = {Physical Review}, 
  issn     = {0031-899X}, 
  doi      = {10.1103/physrev.58.1098}, 
  pages    = {1098--1113}, 
  number   = {12}, 
  volume   = {58}
}

@misc{sm,
  author       = {Tanaya Haldar and Ashis K. Nandy and Anamitra Mukherjee},
  title        = {Supplemental Material},
  year         = {2026},
  note         = {}
}

\clearpage
\onecolumngrid

\setcounter{equation}{0}
\renewcommand{\theequation}{S\arabic{equation}}
\setcounter{figure}{0}
\renewcommand{\thefigure}{S\arabic{figure}}
\setcounter{table}{0}
\renewcommand{\thetable}{S\arabic{table}}

\begin{center}
{\large\bfseries Supplemental Material for\\[0.3em]
``Entropy-Driven Altermagnetism from Thermal Magnons''\par}
\vspace{0.8em}
{\normalsize Tanaya Halder, Ashis K. Nandy, and Anamitra Mukherjee\par}
\vspace{0.35em}
{\small
School of Physical Sciences, National Institute of Science Education and Research,
Jatni 752050, India\\
Homi Bhabha National Institute, Training School Complex, Anushakti Nagar,
Mumbai 400094, India\par}
\end{center}

\begin{quote}
\small
\textbf{Abstract.---}
We provide supporting derivations and numerical details for the
entropy-driven altermagnetic mechanism discussed in the main text.
We derive the $d_{xy}$ exchange modulation and the resulting
sign-changing magnon splitting, establish the thermal softening of the
exchange coordinate and its entropy gain within spin-wave theory, and
give the corresponding spin-integrated description
$K_{\rm eff}=K-\chi_{\rm nem}^{(0)}$ for the classical model.
We also present the Monte Carlo finite-size analysis and the transverse
spin-conductivity calculation, including its low-temperature activated
form and averaging over exchange-modulation fluctuations.
\end{quote}
\normalsize

\section{S1. $d_{xy}$ exchange coordinate}

\paragraph{\textbullet~Exchange pattern and symmetry.---}
The diagonal exchange in the main text is
$J_{2,i\mu}=J_2+\sigma_i\eta_\mu\delta$, with
$\mathbf d_{1,2}=(1,\pm1)$ and $\eta_1=-\eta_2=1$.
The modulation therefore enters the spin Hamiltonian as
\begin{equation}
H_\delta
=
\delta\sum_{i,\mu}\sigma_i\eta_\mu\,
\mathbf S_i\!\cdot\!\mathbf S_{i+\mathbf d_\mu}
=
-\delta\,\mathcal O_{\rm nem},
\label{S:Hdelta}
\end{equation}
where $\mathcal O_{\rm nem}$ is the bond-nematic operator introduced in the
main text. The corresponding diagonal form factor is
\begin{equation}
\sum_{\mu}\eta_\mu
\cos(\mathbf k\!\cdot\!\mathbf d_\mu)
=
\cos(k_x+k_y)-\cos(k_x-k_y)
=
-2\sin k_x\sin k_y .
\label{S:dxyform}
\end{equation}
Thus the exchange modulation transforms as $d_{xy}$: it changes sign under
$C_4$ and distinguishes the two diagonal directions. The two signs of
$\delta$ correspond to the two symmetry-related bond-nematic domains.

For the collinear N\'eel state, NNN spins are parallel and the modulation
does not change the classical energy:
\begin{equation}
E_{\delta}^{\rm cl}
=
S^2\delta\sum_{i,\mu}\sigma_i\eta_\mu
=
S^2\delta\sum_i\sigma_i(\eta_1+\eta_2)
=
0 .
\label{S:Eclassdelta}
\end{equation}
Hence the classical N\'eel state does not favor finite $\delta$; for
$K,U>0$ the bare exchange-mode energy selects $\delta=0$ at zero temperature.

\paragraph{\textbullet\ Microscopic realization.---}
The exchange coordinate $\delta$ can arise from a microscopic lattice
displacement or other soft mode that modulates the diagonal superexchange. For a symmetry-compatible soft mode
$Q$ that modulates the diagonal exchange linearly, $\delta=gQ$, a bare
potential
\begin{equation}
V(Q)=\frac{K_Q}{2}Q^2+\frac{U_Q}{4}Q^4
\end{equation}
maps onto the form used in the main text with
$K=K_Q/g^2$ and $U=U_Q/g^4$, $g$ being the coupling constant as discussed below. A soft phonon coupled to superexchange provides
one microscopic realization, while the minimal theory only requires a
sufficiently soft exchange coordinate in the same $d_{xy}$ symmetry channel.


For a structural realization of the exchange coordinate in Eq.~(1), let
$Q$ denote the displacement amplitude of an internal mode, with a fixed
normalization, whose linear exchange response has the required staggered
pattern. Restoring energy units, we write
\begin{equation}
 J_{2,i\mu}(Q)=J_2+\sigma_i\eta_\mu gQ+O(Q^2),
 \qquad \delta_{\rm phys}=gQ,
 \label{eq:s8_displacement}
\end{equation}
where $g$ is the exchange derivative projected onto that mode and has
units of energy per length. It includes changes of the relevant bond
lengths and angles; a uniform shear lacking the sublattice factor
$\sigma_i$ does not realize the exchange pattern assumed here.
For a bare elastic energy per spin
$f_{\rm el}(Q)=\kappa_Q Q^2/2+u_Q Q^4/4$, the dimensionless coefficients
used in the simulations are
\begin{equation}
 \bar K=\frac{\kappa_Q J_1}{g^2},\qquad
 \bar U=\frac{u_Q J_1^3}{g^4},\qquad
 x\equiv\frac{\delta_{\rm phys}}{J_1}=\frac{gQ}{J_1},
 \label{eq:s8_elastic}
\end{equation}
with $\bar K=0.1$ and $\bar U=15$ for the parameters studied.

As an illustrative scale conversion, matching the model's
$k_BT_N/J_1\simeq1.75$ to $T_N=100$--$300\,\mathrm K$ gives
$J_1\simeq5$--$15\,\mathrm{meV}$, retaining the same spin normalization.
A representative modulation $x=0.1$ then corresponds to
$\delta_{\rm phys}\simeq0.5$--$1.5\,\mathrm{meV}$ and
\begin{equation}
 |Q|\simeq(0.01\text{--}0.03)\,\text{\AA}\,
 \frac{50\,\mathrm{meV}/\text{\AA}}{|g|}.
 \label{eq:s8_scale}
\end{equation}
The value $50\,\mathrm{meV}/\text{\AA}$ is an illustrative benchmark, not a
material-specific determination. Both $g$ and the mode stiffness must
be independently established for a candidate material.
The harmonic magnon energy splitting is
\begin{equation}
 \Delta E_{\mathbf k}=8S\delta_{\rm phys}\sin k_x\sin k_y,
 \qquad \max_{\mathbf k}|\Delta E_{\mathbf k}|=8S|\delta_{\rm phys}|,
 \label{eq:s8_splitting}
\end{equation}
yielding a maximum splitting of approximately $(4\text{--}12)S\,\mathrm{meV}$
for this example, within the locally stable spin-wave regime.
These conditional estimates connect the model to displacement and
spectroscopic scales; they are not a microscopic fit to a specific
compound. The distinctive opportunity is the thermally controlled
emergence of altermagnetic splitting within an ordered antiferromagnet.

\section{S2. Linear spin-wave spectrum and altermagnetic branch reversal}

\paragraph{\textbullet\ Harmonic spin-wave Hamiltonian.---}
We expand about the collinear N\'eel state, with $A\uparrow$ and
$B\downarrow$, using
\begin{align}
A:\quad&
S_i^z=S-a_i^\dagger a_i,
\qquad
S_i^+\simeq\sqrt{2S}\,a_i,
\\
B:\quad&
S_j^z=-S+b_j^\dagger b_j,
\qquad
S_j^+\simeq\sqrt{2S}\,b_j^\dagger .
\end{align}
For the $N/2$ momenta in the magnetic Brillouin zone (MBZ),
$|k_x+k_y|\leq\pi$ and $|k_x-k_y|\leq\pi$, the quadratic Hamiltonian
separates into two conjugate bosonic BdG blocks. In the basis
$\Psi_{\bk}=(a_{\bk},b_{-\bk}^\dagger)^T$, one block is
\begin{equation}
{\cal H}_{\bk}
=
\begin{pmatrix}
\Omega_{\bk}-\lambda_{\bk}\delta & B_{\bk}\\
B_{\bk} & \Omega_{\bk}+\lambda_{\bk}\delta
\end{pmatrix},
\label{S:Hk}
\end{equation}
with
\begin{equation}
\Omega_{\bk}
=
4J_1S+2DS-4J_2S(1-\cos k_x\cos k_y),
\qquad
B_{\bk}
=
2J_1S(\cos k_x+\cos k_y),
\qquad
\lambda_{\bk}
=
4S\sin k_x\sin k_y .
\label{S:OBL}
\end{equation}

The bosonic eigenvalue problem
$\det({\cal H}_{\bk}-\omega\sigma_z)=0$ gives the positive-frequency
mode $R_{\bk}-\lambda_{\bk}\delta$, where
\begin{equation}
R_{\bk}
=
\sqrt{\Omega_{\bk}^2-B_{\bk}^2}.
\end{equation}
The conjugate block gives $R_{\bk}+\lambda_{\bk}\delta$. Thus the two
opposite-spin magnon branches are
\begin{equation}
\omega_\pm(\bk,\delta)
=
R_{\bk}\pm\lambda_{\bk}\delta,
\qquad
\lambda_{\bk}=4S\sin k_x\sin k_y .
\label{S:disp}
\end{equation}
The collinear N\'eel state is locally stable provided
$R_{\bk}>|\lambda_{\bk}\delta|$ throughout the MBZ.

\paragraph{\textbullet\ Eigenvectors and branch reversal.---}
The exchange modulation shifts the magnon energies without changing their
Bogoliubov eigenvectors. Indeed,
\begin{equation}
\sigma_z{\cal H}_{\bk}
=
-\lambda_{\bk}\delta\,\mathbbm{1}_{2\times2}
+
\begin{pmatrix}
\Omega_{\bk} & B_{\bk}\\
-B_{\bk} & -\Omega_{\bk}
\end{pmatrix},
\label{S:identity}
\end{equation}
so the $\delta$-dependent term is proportional to the identity in the
dynamical matrix. The eigenvectors therefore depend only on
$\Omega_{\bk}$ and $B_{\bk}$, while the branch splitting is
\begin{equation}
\omega_+(\bk,\delta)-\omega_-(\bk,\delta)
=
2\lambda_{\bk}\delta
=
8S\delta\sin k_x\sin k_y .
\label{S:splitting}
\end{equation}

Since $\lambda_{\bk}\propto\sin k_x\sin k_y$ has $d_{xy}$ symmetry, the
splitting changes sign under a $C_4$ rotation. Consequently, the two
opposite-spin branches reverse their ordering between
$\Gamma\rightarrow M_+=(\pi/2,\pi/2)$ and
$\Gamma\rightarrow M_-=(\pi/2,-\pi/2)$, as shown in
Fig.~\ref{fig:variational}(c,d). This branch reversal is the magnon
signature of the altermagnetic state.

\section{S3. Variational free energy and spin-wave thermodynamics}

\paragraph{\textbullet\ Nematic softening of the exchange coordinate.---}
For the $d_{xy}$ exchange pattern of Sec.~S1, the classical N\'eel energy is
independent of $\delta$. The harmonic zero-point contribution is also
$\delta$ independent because
$\omega_+(\bk,\delta)+\omega_-(\bk,\delta)=2R_{\bk}$.
Thus, all $\delta$ dependence of the magnetic free energy within LSWT is
thermal, and the equilibrium modulation $\delta_\star(T)$ is obtained by
minimizing Eq.~\eqref{eq:fullfree} using the full magnon dispersion over the
MBZ. No expansion about a particular momentum or dilute-magnon approximation
is used in this minimization.

The thermal softening can be expressed through the $d_{xy}$ bond-nematic
susceptibility associated with the operator $\mathcal O_{\rm nem}$ defined in
Sec.~S1:
\begin{equation}
\chi_{\rm nem}(T)
=
-\left.
\frac{\partial^2 F_{\rm spin}(\delta,T)}
{\partial\delta^2}
\right|_{\delta=0}
=
\frac{\beta}{N}
\left(
\langle\mathcal O_{\rm nem}^2\rangle
-\langle\mathcal O_{\rm nem}\rangle^2
\right),
\label{S:chiNemDef}
\end{equation}
where $F_{\rm spin}$ denotes the spin free energy per site. Using the
spin-wave spectrum of Sec.~S2 gives
\begin{equation}
\chi_{\rm nem}^{\rm SW}(T)
=
\frac{2}{T}
\int_{\rm MBZ}\frac{d^2k}{(2\pi)^2}\,
\lambda_{\bk}^2
n_B(R_{\bk})\bigl[1+n_B(R_{\bk})\bigr].
\label{S:chiNemSW}
\end{equation}
Consequently,
\begin{equation}
K_{\rm eff}(T)=K-\chi_{\rm nem}^{\rm SW}(T).
\label{S:Keff}
\end{equation}
Since $\chi_{\rm nem}^{\rm SW}(0)=0$, the unmodulated state is stable at zero
temperature. Thermal population enhances the nematic fluctuations, and the
quadratic instability occurs when
$\chi_{\rm nem}^{\rm SW}(T)>K$. Higher-order thermal terms also renormalize
the quartic coefficient; $U$ is chosen such that the total quartic term
remains positive over the temperature range considered.

\paragraph{\textbullet\ Energy--entropy balance.---}
To expose the thermodynamic origin of the instability, we decompose the
free-energy difference at the variational minimum as
\begin{equation}
\Delta f(T)
=
F(\delta_\star,T)-F(0,T)
=
\Delta e(T)-T\Delta s(T),
\label{S:DeltaF}
\end{equation}
where
\begin{equation}
\begin{split}
\Delta e(T)
={}&
\frac{K}{2}\delta_\star^2+\frac{U}{4}\delta_\star^4
\\
&+
\int_{\rm MBZ}\frac{d^2k}{(2\pi)^2}
\sum_{s=\pm}
\left\{
\bigl[\omega_s n_B(\omega_s)\bigr]_{\delta_\star}
-
\bigl[\omega_s n_B(\omega_s)\bigr]_{0}
\right\},
\end{split}
\label{S:DeltaE}
\end{equation}
and
\begin{equation}
\Delta s(T)
=
s_{\rm mag}(\delta_\star,T)-s_{\rm mag}(0,T).
\label{S:DeltaS}
\end{equation}
The magnon entropy is
\begin{equation}
s_{\rm mag}(\delta,T)
=
\int_{\rm MBZ}\frac{d^2k}{(2\pi)^2}
\sum_{s=\pm}
\left[
(1+n_s)\ln(1+n_s)-n_s\ln n_s
\right],
\label{S:Smag}
\end{equation}
with $n_s=n_B[\omega_s(\bk,\delta)]$.
The classical and zero-point contributions cancel from these differences
because they are independent of $\delta$. As shown in
Fig.~\ref{fig2}(b), the finite-$\delta$ state initially carries a positive
energetic cost, but its larger magnon entropy eventually gives
$T\Delta s>\Delta e$, so that $\Delta f<0$. This directly identifies the
finite-temperature instability as entropy driven.

\paragraph{\textbullet\ Staggered moment and range of validity.---}
As shown in Sec.~S2, the $\delta$-dependent term is proportional to the
identity in the bosonic dynamical matrix and therefore does not modify the
Bogoliubov eigenvectors. Using
$u_{\bk}^2+v_{\bk}^2=\Omega_{\bk}/R_{\bk}$, the staggered moment is
\begin{equation}
m_s(\delta,T)
=
S-
\int_{\rm MBZ}\frac{d^2k}{(2\pi)^2}
\left\{
\frac{\Omega_{\bk}}{R_{\bk}}
\left[
1+n_B(\omega_+)+n_B(\omega_-)
\right]
-1
\right\}.
\label{S:ms}
\end{equation}
The result shown in the inset of Fig.~\ref{fig2}(a) is evaluated along the
variational minimum,
$m_s(T)=m_s[\delta_\star(T),T]$. The onset of finite
$\delta_\star$ occurs while $m_s$ remains finite, confirming within LSWT
that the exchange modulation develops inside the N\'eel phase.


\section{S4. Transverse spin conductivity and low-temperature limit}

\paragraph{\textbullet\ Linear-$\delta$ response.---}
Starting from Eq.~\eqref{eq:sigmafull}, we write
\[
\omega_s(\bk,\delta)=R_{\bk}+s\lambda_{\bk}\delta,
\qquad
v_{s,a}=v_a+s\delta\lambda_a,
\]
where
$v_a=\partial_{k_a}R_{\bk}$ and
$\lambda_a=\partial_{k_a}\lambda_{\bk}$.
Expanding the conductivity to linear order in $\delta$ gives
\begin{equation}
\frac{\sigma_{xy}^{z}}{\tau}
=
2\delta
\int_{\rm MBZ}\frac{d^2k}{(2\pi)^2}
\left[
(v_x\lambda_y+v_y\lambda_x)\mathcal D
+\lambda_{\bk}v_xv_y\mathcal D'
\right]
+O(\delta^3),
\label{S:siglinear}
\end{equation}
with
$\mathcal D=\partial_T n_B(R_{\bk})$ and
$\mathcal D'=\partial_{R_{\bk}}\mathcal D$.
Thus, the transverse response vanishes for $\delta=0$ and changes sign under
$\delta\rightarrow-\delta$.

\paragraph{\textbullet\ Low-temperature activated form.---}
For $T\ll\Delta_X$, the transport integral is dominated by magnons near
$X=(\pi,0)$. Writing $\bk=X+\bq$,
\begin{equation}
R_{X+\bq}\simeq
\Delta_X+2J_2S q^2,
\qquad
\lambda_{X+\bq}\simeq-4S q_xq_y,
\qquad
\Delta_X=4J_1S+2DS-8J_2S .
\label{S:Xexpansion}
\end{equation}
Since $\lambda_X=0$, $\Delta_X$ is the common magnon gap; the altermagnetic
modulation enters through the momentum dependence of $\lambda_{\bk}$ around
the minimum. In the activated regime,
$n_B(R_{\bk})\simeq e^{-R_{\bk}/T}$, so Eq.~\eqref{S:siglinear} can be
evaluated analytically, yielding

\begin{equation}
\frac{\sigma_{xy}^{z}}{\tau}
\simeq
-\frac{\delta}{\pi J_2}
(\Delta_X+2T)e^{-\Delta_X/T}.
\label{S:sigmalowT}
\end{equation}
The exponential factor reflects activation across the magnon gap, while the
factor of $\delta$ gives the sign change under $\delta\rightarrow-\delta$.
All numerical results in Fig.~\ref{fig2}(d) are obtained instead from the
full-band expression, Eq.~\eqref{eq:sigmafull}, using the equilibrium
$\delta_\star(T)$ from the variational free energy.

\section{S5. Classical spin--lattice Monte Carlo}
\label{sec:SM_MC}

\paragraph{\textbullet\ Observables and susceptibilities.---}
We simulate Eq.~\eqref{eq:H} with classical spins of fixed length
$|\mathbf S_i|=S$, sampling the spin configurations together with the signed
global exchange coordinate $\delta$. The $d_{xy}$ bond-nematic variable is
defined from the operator introduced in Sec.~S1 as
\[
\psi=-\frac{\mathcal O_{\rm nem}}{N},
\]
Because the
finite system samples both symmetry-related signs, the quantities displayed
in Fig.~\ref{fig:f3}(a) are $\langle|\delta|\rangle$ and
$\langle|\psi|\rangle$.

Magnetic order is characterized by the static spin structure factor
\begin{equation}
S(\mathbf q)=
\frac{1}{N}
\sum_{ij}
e^{i\mathbf q\cdot(\mathbf r_i-\mathbf r_j)}
\left\langle
\mathbf S_i\!\cdot\!\mathbf S_j
\right\rangle ,
\qquad
M^2=\frac{S(\pi,\pi)}{N}.
\label{S:MC_structure}
\end{equation}
The finite-size magnetic and exchange-mode susceptibilities are
\begin{equation}
\chi_M=
\frac{N}{T}
\left(
\langle M^2\rangle-\langle M\rangle^2
\right),
\qquad
\chi_\delta=
\frac{N}{T}
\left(
\langle\delta^2\rangle-\langle\delta\rangle^2
\right).
\label{S:MC_susc}
\end{equation}
The peaks of $\chi_M$ and $\chi_\delta$ determine $T_N(L)$ and
$T_\delta(L)$, respectively. As a cross-check, we also used the alternative
estimator
$\frac{N}{T}\left(\langle\delta^2\rangle-\langle|\delta|\rangle^2\right)$,
which gives essentially the same $T_\delta(L)$.

To evaluate directly the spin response responsible for softening the exchange
coordinate, we also perform simulations with $\delta$ set to zero. We
define the corresponding bond-nematic susceptibility as
\begin{equation}
\chi_{\rm nem}^{(0)}(T,L)
=
\frac{\beta}{N}
\left[
\langle\mathcal O_{\rm nem}^{\,2}\rangle_0
-
\langle\mathcal O_{\rm nem}\rangle_0^2
\right],
\label{S:MC_chinem0}
\end{equation}
where $\langle\cdots\rangle_0$ denotes the equilibrium ensemble with
$\delta=0$. This is the susceptibility entering the stiffness
renormalization derived in Sec.~S6.

\paragraph{\textbullet\ Monte Carlo protocol and finite-size analysis.---}
Periodic $L\times L$ lattices with $L\leq200$ are simulated using Metropolis
updates of the spins and of the global coordinate $\delta$. Both signs of
$\delta$ are sampled. After equilibration, $5\times10^4$ measurement sweeps
are accumulated at each temperature, with independent runs used to estimate
statistical uncertainties.

The finite-size transition temperatures are extrapolated linearly in $1/L$.
For the coherent exchange mode the scaling variable is
$(T-T_\delta)N^{1/2}$, giving
$T_\delta(L)-T_\delta\propto L^{-1}$ for $N=L^2$, while the easy-axis
magnetic transition has the same leading shift because
$\nu_{\rm Ising}=1$ in two dimensions (Sec.~S6).

\paragraph{\textbullet\ Transport from the Monte Carlo exchange distribution.---}
The Monte Carlo dynamics is used only to obtain the equilibrium distribution
of the exchange coordinate. Since
$\sigma_{xy}^{z}(T,-\delta)=-\sigma_{xy}^{z}(T,\delta)$, the two
symmetry-related sectors are aligned by $\delta\rightarrow|\delta|$.
Denoting the resulting normalized distribution by $P_T(\delta)$ for
$\delta\geq0$, the conductivity shown in Fig.~\ref{fig:f3}(b) is
\begin{equation}
\left\langle
\sigma_{xy}^{z}\right\rangle_{\rm MC}/{\tau}=
\int_0^\infty d\delta\,
P_T(\delta)
\frac{\sigma_{xy}^{z}(T,\delta)}{\tau}.
\label{S:MC_sigma}
\end{equation}
For each sampled $\delta$, $\sigma_{xy}^{z}(T,\delta)/\tau$ is evaluated
from Eq.~\eqref{eq:sigmafull} using the full MBZ magnon spectrum. Thus Monte
Carlo supplies the equilibrium thermal distribution of $\delta$, while LSWT
supplies the transport kernel for each value of $\delta$. This construction
is used only for $T<T_N$, where the N\'eel-state magnon description remains
applicable.

 \paragraph{\textbullet\ Finite-size scaling.---}The finite-size transition temperatures are extrapolated linearly in $1/L$.
For the coherent exchange mode, the scaling form derived in Sec.~S6 gives
$T_\delta(L)-T_\delta\propto N^{-1/2}=L^{-1}$ for $N=L^2$.
The easy-axis magnetic transition has the same leading shift,
$T_N(L)-T_N\propto L^{-1/\nu}=L^{-1}$, since $\nu=1$ for the
two-dimensional Ising universality class.
The magnetic finite-size extrapolation assumes two-dimensional
Ising scaling. Coupling to a relaxing exchange coordinate may
modify the asymptotic critical behavior through an energy-density
feedback analogous to the Larkin--Pikin mechanism
\cite{LarkinPikin1969,ChandraPRResearch2020}.
Our finite-size analysis does not exclude a very weak first-order
magnetic transition; resolving this possibility lies beyond the
present study.

\section{S6. Spin-integrated effective theory and finite-size scaling}
\label{sec:SM_eff}

\paragraph{\textbullet\ Spin-integrated exchange stiffness.---}
For the classical Monte Carlo model, the exchange coordinate enters as
\begin{equation}
H(\delta)=H_0-\delta\mathcal O_{\rm nem}
+N\left(\frac{K}{2}\delta^2+\frac{U}{4}\delta^4\right),
\label{S:Heff_start}
\end{equation}
where $\mathcal O_{\rm nem}$ is the $d_{xy}$ bond operator defined in
Sec.~S1. Integrating out the spins at fixed $\delta$ gives
\begin{equation}
e^{-\beta N F_{\rm eff}(\delta,T)}
\propto
e^{-\beta N(K\delta^2/2+U\delta^4/4)}
\left\langle
e^{\beta\delta\mathcal O_{\rm nem}}
\right\rangle_0 ,
\label{S:Zcumulant}
\end{equation}
where $\langle\cdots\rangle_0$ denotes the unmodulated ($\delta=0$)
spin ensemble. Since $\mathcal O_{\rm nem}$ is odd under $C_4$, its odd
cumulants vanish. To quadratic order,
\begin{equation}
\frac{F_{\rm eff}(\delta,T)}{N}
=
\frac{1}{2}
\left[K-\chi_{\rm nem}^{(0)}(T)\right]\delta^2
+\frac{u_{\rm eff}(T)}{4}\delta^4+O(\delta^6),
\label{S:Feff_cumulant}
\end{equation}
with
\begin{equation}
\chi_{\rm nem}^{(0)}(T)
=
\frac{\beta}{N}
\left[
\langle\mathcal O_{\rm nem}^2\rangle_0
-\langle\mathcal O_{\rm nem}\rangle_0^2
\right].
\label{S:chin0_def}
\end{equation}
Thus
\begin{equation}
K_{\rm eff}(T)=K-\chi_{\rm nem}^{(0)}(T).
\label{S:Keff_chi}
\end{equation}
The superscript $0$ emphasizes that the susceptibility is evaluated with
$\delta$ set to zero distinct from the nematic
susceptibility of the fully coupled spin--$\delta$ system, which becomes
critical together with the exchange mode at the continuous transition.
The quartic coefficient $u_{\rm eff}$ includes the corresponding higher
cumulants and remains positive for the parameters used here.

\paragraph{\textbullet\ Finite-temperature window.---}
At low temperature, the gapped spin-wave population vanishes and hence
$\chi_{\rm nem}^{(0)}\rightarrow0$. At the opposite limit, the variance per
site of the bounded classical bond operator approaches a finite constant,
so
\begin{equation}
\chi_{\rm nem}^{(0)}(T)
\sim \frac{C_\infty}{T}
\longrightarrow0,
\qquad T\rightarrow\infty .
\label{S:chin_highT}
\end{equation}
The exchange-mode softening is therefore restricted to intermediate
temperatures. If
\begin{equation}
\max_T\chi_{\rm nem}^{(0)}(T)>K ,
\label{S:instability_condition}
\end{equation}
$K_{\rm eff}$ changes sign over a finite temperature interval, producing a
reentrant finite-$\delta$ state. Only the part of this interval that
coexists with N\'eel order is altermagnetic.

\paragraph{\textbullet\ Finite-size scaling.---}
Near the continuous exchange-modulation transition, the effective free energy
takes the Landau form
\begin{equation}
F_{\rm eff}\simeq
N\left[
\frac{a(T-T_\delta)}{2}\delta^2+\frac{u}{4}\delta^4
\right],
\qquad u>0 .
\label{S:global_landau}
\end{equation}
Rescaling $\delta=N^{-1/4}\phi$ shows that the scaling variable is
$(T-T_\delta)N^{1/2}$. Hence, the finite-size shift of the susceptibility
maximum obeys
\begin{equation}
T_\delta(L)-T_\delta(\infty)\propto N^{-1/2}=L^{-1},
\qquad N=L^2 .
\label{S:oneoverL}
\end{equation}

\begin{figure*}[t]
 \centering
 \includegraphics[width=\textwidth]{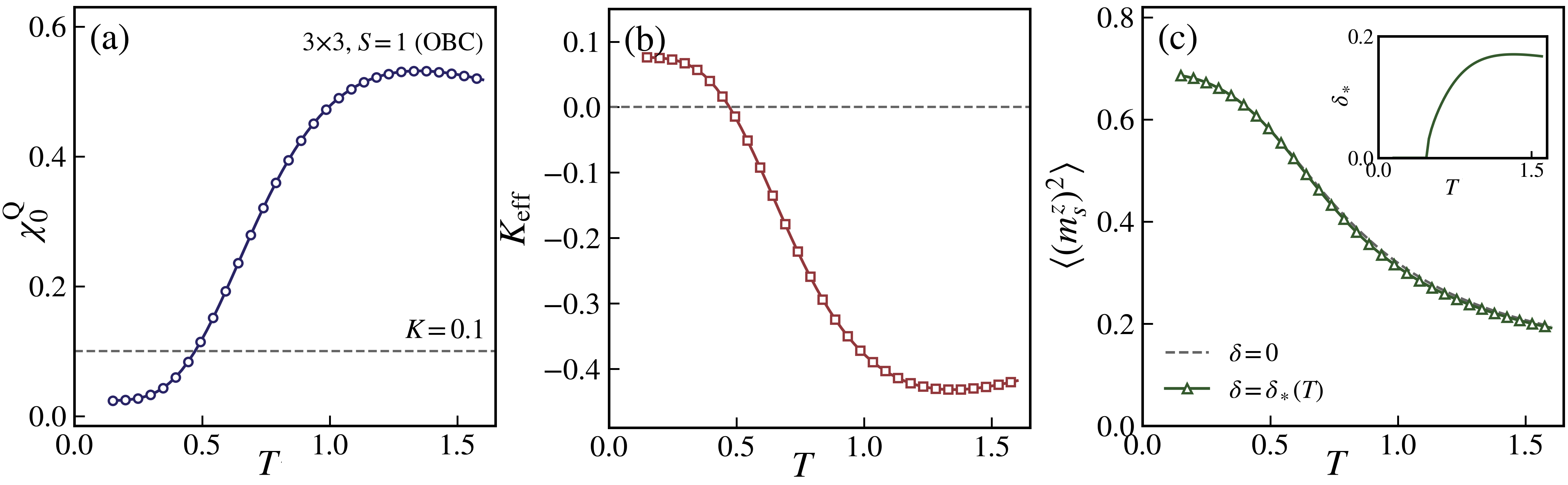}
 \caption{Full-ED quantum exchange softening and equilibrium magnetic correlations for a $3\times3$ open cluster of $S=1$ spins, with $J_1=1$, $J_2=0.35$, $D=0.5$, $K=0.1$, and $U=15$.
 (a) Exchange susceptibility $\chi_0^{\mathrm Q}(T)$ at $\delta=0$; the dashed line marks $K$.
 (b) Effective stiffness $K_{\mathrm{eff}}=K-\chi_0^{\mathrm Q}$; the dashed line marks zero.
 (c) Staggered second moment $\mathcal M_s^2(T)=\langle(m_s^z)^2\rangle_{\delta_\ast(T),T}$ (green triangles), with the fixed-$\delta=0$ result shown as a gray dashed curve. The inset shows the nonnegative equilibrium modulation $\delta_\ast(T)$. All thermal averages include the full spin Hilbert space.}
 \label{fig:s7_ed}
\end{figure*}

\section*{S7. Full exact-diagonalization check of quantum exchange softening}

To test the exchange-softening mechanism of the Hamiltonian in Eq.~(1) of the main text beyond the spin-wave approximation, we perform full exact diagonalization (ED) on a $3\times3$ cluster with open boundaries. We choose $S=1$, the smallest spin for which the single-ion anisotropy is nontrivial: for $S=\tfrac12$, $(S_i^z)^2=\tfrac14$ is a constant, while. for $S\geq2$, the Hilbert space is too large for any meaningful exact diagonalization. The choice of $S=1$ retains the easy-axis physics while keeping the complete Hilbert space, of dimension $3^9=19683$, accessible. We use $J_1=1$, $J_2=0.35$, $D=0.5$, $K=0.1$, and $U=15$, with $k_{\mathrm B}=1$. Open boundaries accommodate the staggered exchange pattern on the odd-sized cluster.

Writing the spin part of the model as
\begin{equation}
 H_{\mathrm{spin}}(\delta)=H_{\mathrm{spin}}(0)-\delta O_{\mathrm{nem}},
 \qquad
 O_{\mathrm{nem}}=-\sum_{i,\mu}\sigma_i\eta_\mu
 \mathbf S_i\cdot\mathbf S_{i+\mathbf d_\mu},
 \label{eq:s7_operator}
\end{equation}
we retain only bonds lying within the cluster. Here $\sigma_i=(-1)^{x_i+y_i}$, $\mathbf d_{1,2}=(1,\pm1)$, and $\eta_{1,2}=\pm1$, as in the main text. At each fixed $\delta$, all eigenvalues $E_n(\delta)$ are obtained in every total-magnetization sector $M=\sum_i S_i^z$. Spin reversal permits the $M>0$ spectra to be counted twice, together with the $M=0$ spectrum counted once. Thus, the partition function includes every state, without a low-energy truncation:
\begin{equation}
 Z(\delta,T)=\sum_n e^{-\beta E_n(\delta)},\qquad
 f_{\mathrm{spin}}(\delta,T)=-\frac{T}{N}\ln Z(\delta,T),
 \quad \beta=T^{-1},\quad N=9.
 \label{eq:s7_partition}
\end{equation}
The sum includes all magnetization sectors and degeneracies. The free energy is per site and includes the quantum ground-state contribution.

The exchange susceptibility and effective stiffness are defined by
\begin{align}
 \chi_{nem,0}^{\mathrm Q}(T)
 &=-\left.\frac{\partial^2 f_{\mathrm{spin}}}{\partial\delta^2}\right|_{\delta=0}
 \simeq-\frac{f_{\mathrm{spin}}(\epsilon,T)+f_{\mathrm{spin}}(-\epsilon,T)
 -2f_{\mathrm{spin}}(0,T)}{\epsilon^2},\nonumber\\
 K_{\mathrm{eff}}(T)&=K-\chi_{nem,0}^{\mathrm Q}(T).
 \label{eq:s7_curvature}
\end{align}
We use $\epsilon=0.0125$, with $\epsilon=0.025$ as a finite-difference check. The equilibrium modulation is the nonnegative free-energy minimum,
\begin{equation}
 \delta_\ast(T)=\underset{\delta\geq0}{\operatorname{arg\,min}}
 \left[f_{\mathrm{spin}}(\delta,T)+\frac K2\delta^2+\frac U4\delta^4\right].
 \label{eq:s7_minimum}
\end{equation}
The two signs of $\delta_\ast$ are symmetry related. As in the variational calculation, $\delta$ is treated as a static coordinate. We minimize an interpolation in $\delta^2$ of the ED free energies, using a modulation spacing of $0.0125$ in the region containing the minima; grid refinement and direct diagonalizations at selected minima were employed to verify the interpolation.

To assess the magnetic correlations along this equilibrium branch, we calculate
\begin{equation}
 \mathcal M_s^2(T)\equiv
 \left\langle(m_s^z)^2\right\rangle_{\delta_\ast(T),T}
 =\left.\frac{\sum_n e^{-\beta E_n(\delta)}
 \langle n,\delta|(m_s^z)^2|n,\delta\rangle}{Z(\delta,T)}\right|_{\delta=\delta_\ast(T)},
 \qquad
 m_s^z=\frac1N\sum_i\sigma_i S_i^z.
 \label{eq:s7_magnetic}
\end{equation}
The matrix elements are evaluated from the ED eigenvectors. This second moment is the finite-cluster magnetic diagnostic corresponding to the moment evaluated at $\delta_\ast(T)$ in the inset of Fig.~2(a) of the main text.

Figure~\ref{fig:s7_ed} shows that $\chi_0^{\mathrm Q}$ exceeds $K$ on heating near $T/J_1\simeq0.47$, where the unmodulated minimum loses local stability. A finite $\delta_\ast$ develops while substantial staggered correlations persist along the minimizing branch. For example, at $T/J_1\simeq1.01$, $\delta_\ast/J_1\simeq0.162$ and $\mathcal M_s^2\simeq0.308$. These results support quantum exchange softening in a strongly antiferromagnetically correlated finite cluster. They do not determine a thermodynamic magnetic ordering temperature or establish long-range altermagnetic order.

\end{document}